\documentclass[aip, reprint, graphicx]{revtex4-1}
\usepackage[version=3]{mhchem}
\usepackage{mathtools}
\usepackage{setspace}
\usepackage{dcolumn}
\usepackage{bm}

\usepackage{float}
\usepackage{array}
\usepackage{sidecap}
\usepackage{epstopdf}
\newcolumntype{L}[1]{>{\raggedright\let\newline\\\arraybackslash\hspace{0pt}}m{#1}}
\newcolumntype{C}[1]{>{\centering\let\newline\\\arraybackslash\hspace{0pt}}m{#1}}
\newcolumntype{R}[1]{>{\raggedleft\let\newline\\\arraybackslash\hspace{0pt}}m{#1}}

\begin{document}

\title{Multi-mode technique for the determination of the biaxial \ce{Y2SiO5} permittivity tensor from 300 to 6 Kelvin} 

\author{N. C. Carvalho}
\email[]{natalia.docarmocarvalho@research.uwa.edu.au}
\affiliation{School of Physics, The University of Western Australia, Crawley, 6009, Australia}
\affiliation{ARC Centre of Excellence for Engineered Quantum Systems (EQuS), 35 Stirling Hwy, 6009, Crawley, Australia }
\author{J-M. Le Floch}
\affiliation{School of Physics, The University of Western Australia, Crawley, 6009, Australia}
\affiliation{ARC Centre of Excellence for Engineered Quantum Systems (EQuS), 35 Stirling Hwy, 6009, Crawley, Australia }
\author{J. Krupka}
\affiliation{Instytut Mikroelektroniki i Optoelektroniki PW, Koszykowa 75, 00-662 Warsaw, Poland}
\author{M. E. Tobar}
\affiliation{School of Physics, The University of Western Australia, Crawley, 6009, Australia}
\affiliation{ARC Centre of Excellence for Engineered Quantum Systems (EQuS), 35 Stirling Hwy, 6009, Crawley, Australia }

\date{\today}

\begin{abstract}
The \ce{Y2SiO5} (YSO) crystal is a dielectric material with biaxial anisotropy with known values of refractive index at optical frequencies. It is a well-known rare-earth (RE) host material for optical research and more recently has shown promising performance for quantum-engineered devices. In this paper, we report the first microwave characterization of the real permittivity tensor of a bulk YSO sample, as well as an investigation of the temperature dependence of the tensor components from 296 K down to 6 K. Estimated uncertainties were below 0.26\%, limited by the precision of machining the cylindrical dielectric. Also, the electrical Q-factors of a few electromagnetic modes were recorded as a way to provide some information about the crystal losses over the temperature range. To solve the tensor components necessary for a biaxial crystal, we developed the multi-mode technique, which uses simultaneous measurement of low order Whispering Gallery Modes. Knowledge of the permittivity tensor offers important data, essential for the design of technologies involving YSO, such as microwave coupling to electron and hyperfine transitions in RE doped samples at low temperatures.
\end{abstract}

\maketitle 

Yttrium orthosilicate (\ce{Y2SiO5} or YSO) is a low loss dielectric solid material especially interesting when doped with rare-earth (RE) ions.  It has proved to be a great host due the small magnetic moments of its constituents, allowing long dephasing times of the dopant ions, including a record 6 hours coherence of nuclear spins in $^{151}$Eu$^{3+}:$\ce{Y2SiO5}\cite{Sellars}. Also, as a transparent crystal it can substitute atomic gases in several applications, being advantageous due the high density of atoms, compactness and by having no atomic diffusion \cite{ham1997, wang2007}.
Such characteristics led researchers to use RE doped YSO crystals for a variety of purposes. For the last few decades the material has been used as an excellent laser \cite{li1992} and stable frequency source\cite{nist1,nist2}, and its luminescence properties have been thoroughly investigated \cite{shin2001, Ramakrishna2014, lin1996, wen2014}. The scintillation effect of the YSO doped crystals made them also interesting for image processing \cite{zorenko2014, yamamoto2014}, including applications with biological purposes \cite{taylor2013, hsu2014, Thibault2006}. More recently, the growing interest in quantum information science has turned the focus of YSO \cite{rielander2014, lauritzen2010, sabooni2013} to the microwave domain, which has showed being valuable for quantum information processing through optical \cite{Sellars, fraval2005, zheng2014} and hybrid devices \cite{probst2014, goryachev2014}. 

\begin{equation}
	\begin{pmatrix}
		\epsilon_x &0 & 0 \\
  		0 & \epsilon_y & 0 \\
  		0 & 0 & \epsilon_z 
 	\end{pmatrix}
	\label{Eq1}
\end{equation}

YSO is a monoclinic biaxial crystal, which means the real part of its complex permittivity can always be diagonalized and then defined by three different values distributed along the main diagonal of a second rank tensor, and by ignoring the loss terms, the tensor is is given by Eq. (\ref{Eq1})\cite{petit2010, jellison2011}.

Since the crystal has been used extensively in the optical, the components along the principal axes are well known at these frequencies and follow a Sellmeier dispersion equation \cite{beach1990}. However, the permittivity of such type of crystals has not been fully determined so far at microwave frequencies, at which the aforementioned optical equations are invalid. In this paper we present a technique to determine the real part of the YSO permittivity tensor along with an estimate of the dielectric loss in the microwave range. This study also investigates the temperature dependence of these properties through what we call the multi-mode technique, which is based on the Whispering Gallery Mode (WGM) technique \cite{krupka19992, tobar2001}, used for the characterization of isotropic and uniaxial crystals. This method likewise uses WGMs to perform the permittivity characterization. However, necessarily, also uses lower order mode frequencies in the experiment. 

\begin{figure}
\begin{center}
	\includegraphics[height=2.8cm]{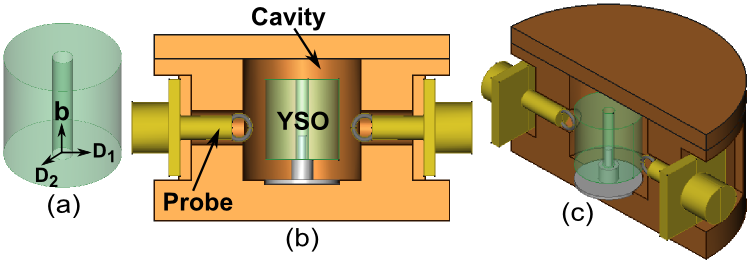}
	\caption{(a) Diagram of the YSO cylindrical sample and its alignment to the three principle axes of polarization. (b) Schematic representation of the microwave cavity cross-section. (c) Axonometric view of the cavity scheme.\label{fig1}}
\end{center}
\end{figure}

\begin{figure*}
\begin{center}
	\includegraphics[height=4.8cm]{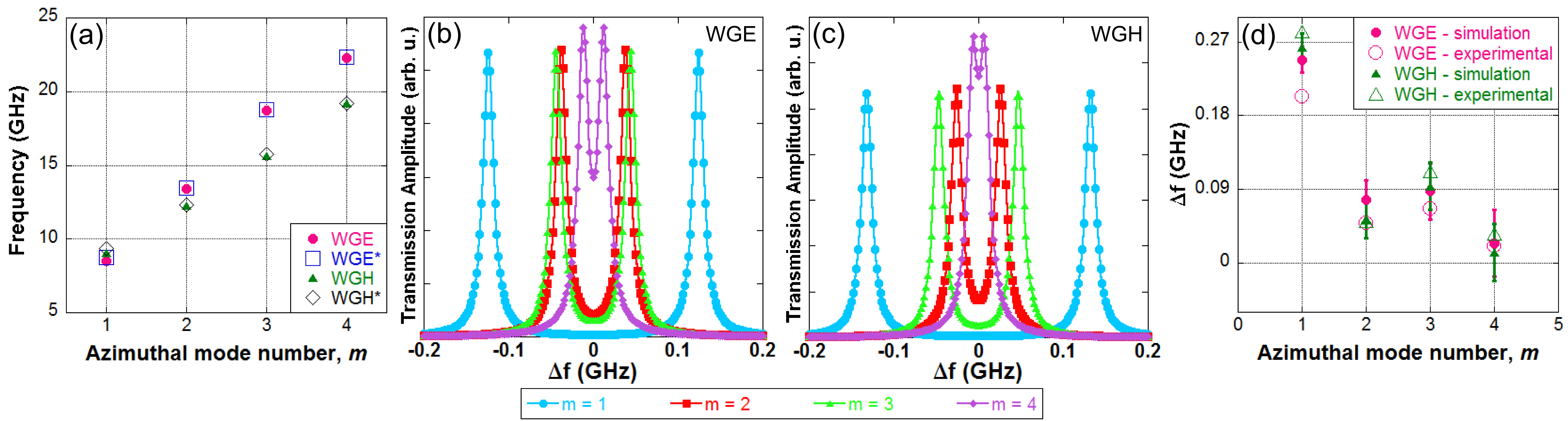}
	\caption{Room temperature measurements: (a) Measured mode frequency versus azimuthal mode number (\textit{m}). Doublets pairs are marked with an asterisk. (b) WGE$_{m11}$ and (c) WGH$_{m11}$ mode splitting of the doublets pairs fitted by a Lorentzian function with experimental frequency separation (\textit{$\Delta$f}) and arbitraries FWHW (full width at half maximum) and amplitude. (d) Frequency separation of the doublets pairs as a function of the azimuthal mode number. \label{fig2}}
\end{center}
\end{figure*}

The Split-post Dielectric Resonator (SPDR) also is a commonly used technique in dielectric metrology. It allows the measurement of the electric permittivity and loss tangent of flat laminar dielectric specimens in the frequency range of 1 - 10 GHz \cite{krupka2001, lefloch2014}. In this work, such a technique was employed to perform an initial relatively inaccurate measurement of the real permittivity tensor of the YSO with three substrates  in which the thickness dimension varied between 0.4 to 0.5 mm between samples with 2\% variation within one sample. The flat laminar size of the substrates was approximately 10 mm by 10 mm, with each sample having one of the principal axes along the thickness dimension. Thus, excitation of the transverse electric mode in the SPDR meant that the direction of the electric field (E-field) was perpendicular to the thickness direction, so the E-field sampled the two principal axes perpendicular to the thickness. Thereby, the three separate substrate measurements were used to estimate the geometric average of the two perpendicular principle axes, then with three equations and three unknowns, we could estimate the permittivity tensor as: $\epsilon_x$ = 9.6 $\pm$ 0.2, $\epsilon_y$ = 11.0 $\pm$ 0.2 and $\epsilon_z$ = 10.3 $\pm$ 0.2.

Nevertheless, this method is not adequate to provide accurate results for biaxial samples due to the imprecision in the substrate thickness, so we refined these values by developing the multi-mode technique for a well-defined bulk cylinder. The right cylinder diameter and height were measured to be 10.004  $\pm$ 0.009 mm and 9.829  $\pm$ 0.009 mm, respectively. Also, the sample had a concentric hole through the middle of the cylinder axis with diameter equal to 1.600 $\pm$ 0.001 mm. The crystal was made of Er$^{3+}$:YSO 0.001\% by Scientific Materials Corp. and it has its b-axis parallel to the cylinder axis Fig. \ref{fig1}(a). The impurities have little effect on the mode frequencies at room temperature with zero field splitting distinct for microwave frequencies less than 5 GHz.

The crystal was placed into a cylindrical copper cavity and supported by a Teflon holder, inserted to keep the crystal nearly at the center of the cavity (Figs. \ref{fig1}(b) and \ref{fig1}(c)). The cavity was designed to enable the two electromagnetic probes to be placed close to the crystal; thus microwave radiation could be injected into the sample exciting its electromagnetic modes.

\begin{table}
		\caption{YSO real permittivity temperature dependence. \label{tab1}}
		\begin{tabular}[c]{ C{2.5cm}  C{1.7cm}  C{1.7cm}  C{1.7cm}} \hline \hline
					
			Temperature  &  $\epsilon_x$ & $\epsilon_y$ & $\epsilon_z$\\ \hline 

			6	&9.36000	&10.90000	&10.21000\\

			10	&9.36000	&10.90000	&10.21000\\

			20	&9.36022	&10.90022	&10.21023\\

			30	&9.36110	&10.90110	&10.21120\\

			40	&9.36300	&10.90300	&10.21310\\

			50	&9.36570	&10.90570	&10.21580\\

			70	&9.37390	&10.91620	&10.22450\\

			296	&9.59750	&11.21750	&10.39000\\  \hline \hline

		\end{tabular}
		\label{tab1}
\end{table}

Several modes were measured between 8 GHz and 22.3 GHz, at these frequencies typically the permittivity remains frequency independent. The loop probes were strategically positioned to couple strongly to the WGM’s, being free to rotate and measure as H$_{\theta}$ as H$_z$. The experiment was performed using a Vector Network Analyzer (VNA) adjusted to operate in transmission, i.e., collecting the scattering parameters S$_{12}$ or S$_{21}$. The YSO resonator is a passive device, therefore S$_{12}$ = S$_{21}$.

\begin{figure*}
	\includegraphics[height=9.2cm]{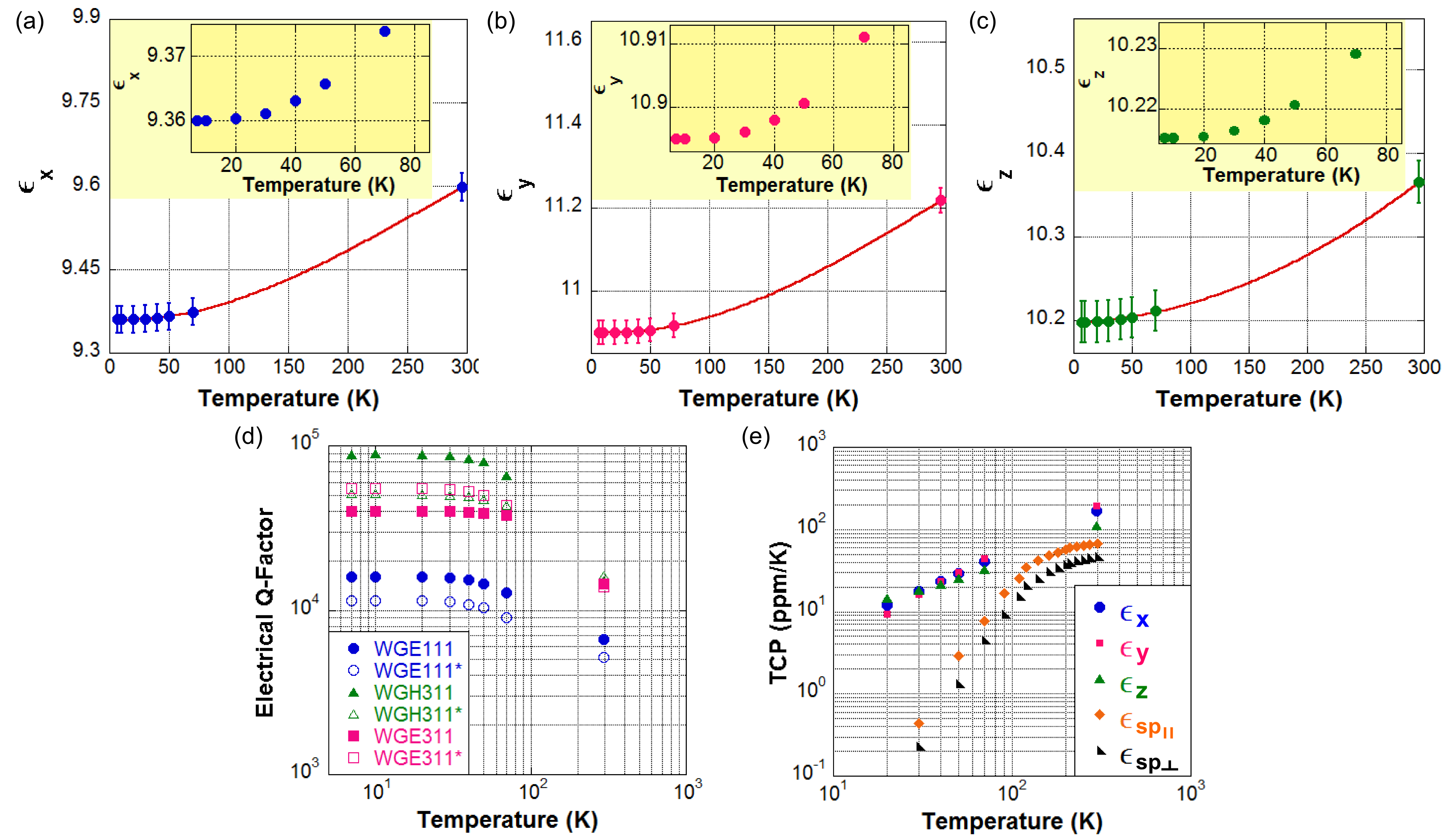}
	\caption{(a) $\epsilon_x$, (b) $\epsilon_y$ and (c) $\epsilon_z$ (YSO permittivity tensor) as a function of temperature. (d) Temperature dependence of the electrical Q-factor of the modes $WGE_{111}$, $WGE_{311}$ and $WGH_{311}$ and their respective doublets pairs (marked with an asterisk). (e) Temperature coefficient of permittivity (TCP) of the YSO ($\epsilon_x$, $\epsilon_y$ and $\epsilon_z$) and Sapphire (sp \textit{parallel} and sp \textit{perpendicular}). \label{fig3}}
\end{figure*}

Finite Element Method (FEM) simulations were used to find out which permittivity tensor would reproduce the experimental observation. The former tensor obtained through the SPDP technique was assumed as a first guess. Then, the aim was matching the experimental mode frequencies measured at room temperature to their simulated counterpart. This was done by varying the permittivity tensor in the model through a sweeping with step $\Delta\epsilon_{i}$ = 0.1, i = {x, y, z}. The mode structures were quantitatively and qualitatively analyzed to label the simulated mode frequencies; the experimental modes were identified handling the probes orientation and position. The WGE$_{m11}$ (Whispering Gallery Electric mode), WGH$_{m11}$ (Whispering Gallery Magnetic mode), TE$_{00}$ (pure Transverse Electric mode) and TM$_{00}$ (pure Transverse Magnetic mode) were the chosen modes to be matched (here, \textit{m} is the azimuthal mode number and went from 1 to 4). The selected WGM’s can be seen on Fig. \ref{fig2}(a).

The WGM technique \cite{krupka1999} states that high order WGMs have to be measured in order to ensure the maximization of mode confinement in the crystal, which must result in a highly accurate permittivity characterization. However, this method had to be adapted to low order WGMs as in biaxial crystals the WGMs appear as pairs, that is, same mode structure at two different frequencies (Figs. \ref{fig2}(b) and \ref{fig2}(c)). The frequency separation depends on the biaxial anisotropy, i.e., the permittivity components in the perpendicular plane to z-axis, $\epsilon_x$  and $\epsilon_y$  (the bigger the difference the bigger the frequency separation). However the separation decreases inversely proportional to the mode frequency as shown by Fig. \ref{fig2}(d), meaning that it is harder to distinguish the anisotropy component of the frequency separation between the two modes in high order WGMs, especially at room temperature, when the electrical quality factor (Q-factor) of the sample is low. On the other hand, the measurement of the frequency separation of the doublet pair is essential to the tensor determination; if the modes are indistinguishable, $\epsilon_x$  and $\epsilon_y$  must be equal and we would have an uniaxial crystal tensor, instead of a biaxial one. Also, we additionally measured the pure transverse electromagnetic modes as a way to certify the consistency of the results. In this sense, we say that the multi-mode technique was applied.

The main diagonal of the real permittivity tensor that best fitted the experiment at room temperature has the following components (see Eq. \ref{Eq1}): $\epsilon_x$ = 9.60 $\pm$ 0.03, $\epsilon_y$ =11.22 $\pm$ 0.03 and $\epsilon_z$ =10.39 $\pm$ 0.03.

This tensor provided that the simulated frequencies reproduced the spectrum of the measured mode frequencies with a discrepancy no higher than 34 MHz (and typically a few MHz). The source of uncertainties was attributed to the irregularities on the crystal dimensions.

Aiming to determine how the permittivity tensor behaves when the material is subject to cryogenic temperatures, the microwave resonant cavity contained the YSO were placed into a cryo-cooler system and cooled down to 6 K. The frequencies of the electromagnetic modes WGE$_{111}$, WGE$_{311}$ and WGH$_{311}$ and their respective doublets pairs were tracked in the warming up process. Then the FEM model revealed how the real permittivity components changed at different temperatures. As the thermal contraction of the copper enclosure was considered in the simulations, the final result presents the temperature dependence of the real permittivity tensor mostly due the crystal internal properties. 

Thus, Figs. \ref{fig3}(a), \ref{fig3}(b) and \ref{fig3}(c) shows the temperature dependence of the three non-zero components of the YSO real permittivity tensor. Measurements were performed from 6 to 297 K, emphasizing the cryogenic temperature range. It can be observed on the plots that the polynomial trend lines fit very well the data points, which supports the consistency of the increment in permittivity with the increasing temperature. Table \ref{tab1} shows the real permittivity non-zero components at the measured temperatures.

The electrical Q-factor of the selected modes was also recorded as a function of temperature. Fig. \ref{fig3}(d) presents the evolution of the Q-factor as the YSO sample warmed up. It can be clearly seen that all modes respect the same trend and the Q-factor is decreasing in the process. Although, this is an expected result, such curves offer a valuable insight of the temperature dependence of the loss terms of the permittivity tensor. The similarities between the curves suggest that the loss terms of the permittivity tensors are close in value, although the lower order modes have less confinement and are more likely to be degraded by metallic losses of the cavity.

Finally, the Fig. \ref{fig3}(e) shows the calculated temperature coefficient of permittivity (TCP) of the YSO. It is only relevant above 20 K as below this temperature typically paramagnetic impurities present in the crystal affect the result. The YSO TCPs are of the same order as other ionic crystals \cite{harrop1969}. Compared to sapphire \cite{krupka19992}, a widely used dielectric with permittivity also around 10, the YSO presents a stronger temperature dependence of permittivity. However, YSO offers the advantage of being a good host material for rare-earth and paramagnetic impurities, due to the smaller effect of the crystal field, allowing narrower spin resonances. Thus, it is a very promising material for quantum hybrid systems.

In summary, this work presented a highly precise determination of the three non-zero components of the YSO real permittivity tensor, which are $\epsilon_x$  = 9.6, $\epsilon_y$ = 11.22 and $\epsilon_z$ = 10.39. The discrepancy between the predicted resonant modes and those were measured was no larger than 34 MHz. The results' uncertainty, associated to the crystal dimensions, is nearly to 0.26\% for $\epsilon_x$, $\epsilon_y$ and $\epsilon_z$. The temperature dependence of the complex permittivity was also investigated, enabling the establishment of a relation between the permittivity components and the YSO temperature in a range from 6 to 296 K. Furthermore, the measurement of the electrical Q-factor at the same temperatures showed an approach to the material losses. In conclusion, the study performed adds a data set to the YSO known properties. There is a high expectation that the YSO may be a key material in quantum information science, moreover, this is already an important crystal for diverse applications. In this sense, the real permittivity tensor measured in this work is important for design and applications at microwave frequencies.

This research is supported by the Australian Research Council CE110001013 and by the Conselho Nacional de Desenvolvimento Cient\'ifico e Tecnol\'ogico (CNPq – Brazil).


%

\end{document}